\documentclass[11pt]{article}

\newtheorem{proposition}{Proposition}
\newtheorem{theorem}{Theorem}
\newtheorem{corollary}{Corollary}

\def\tr{\mathop{\rm tr}\nolimits}

\def\ci{\mathop{\textrm{i}}\nolimits}

\begin{document}

\title{On the Weyl transverse frames in type I spacetimes}
\author{Joan Josep Ferrando$^1$ and Juan Antonio S\'aez$^2$}
\date{\today}

\maketitle

\vspace{2cm}
\begin{abstract}
We apply a covariant and generic procedure to obtain explicit
expressions of the transverse frames that a type I spacetime
admits in terms of an arbitrary initial frame. We also present a
simple and general algorithm to obtain the Weyl scalars
$\Psi_2^T$, $\Psi_0^T$ and $\Psi_4^T$ associated with these
transverse frames. In both cases it is only necessary to choose a
particular root of a cubic expression.
\end{abstract}

\vspace{1cm}

\vspace{2cm}
$^1$ Departament d'Astronomia i Astrof\'{\i}sica, Universitat
de Val\`encia, E-46100 Burjassot, Val\`encia, Spain.\\
E-mail: {\tt joan.ferrando@uv.es}

$^2$ Departament de Matem\`atica Econ\`omico-Empresarial, Universitat de
Val\`encia, E-46071 Val\`encia, Spain\\
E-mail: {\tt juan.a.saez@uv.es}

\newpage

\section{\large INTRODUCTION}

The components $\Psi_a$ of the Weyl tensor in a complex null
tetrad $\{l,k,m, \bar{m}\}$ have a specific physical meaning
\cite{sze}. If an observer lying on the time-like plane $\{l,k\}$
analyzes the deviation of test free particles they can conclude
that the components $\Psi_0$ and $\Psi_4$ describe, respectively,
incoming and outgoing transverse waves, whereas $\Psi_1$ and
$\Psi_3$ are incoming and outgoing longitudinal wave components.
On the other hand, $\Psi_2$ is the Coulomb part of the
gravitational field \cite{sze}.

Depending on the Petrov-Bel type, special frames (and observers)
exist for which the $\Psi_a$ take particular simple forms \cite{sze}
\cite{kra}. Thus, in algebraically general spacetimes we can
consider two different types of adapted frames. If we take the real
null vectors $l$ and $k$ lying on one of the three Weyl principals
planes then $\Psi_1 = \Psi_3 = 0$, the transversal wave components
$\Psi_0$ and $\Psi_4$ being non zero: we have then the {\it Weyl
transverse frames} \cite{bee}. On the other hand, the frames with
$l$ and $k$ aligned with two of the four Debever directions satisfy
$\Psi_0 = \Psi_4 = 0$, the longitudinal wave components $\Psi_1$ and
$\Psi_3$ being non zero: they are the {\it Weyl longitudinal frames}
or {\it Debever frames}.

As a consequence of the peeling-off theorem \cite{sachs}, one of
the transverse components, $\Psi_0$ or $\Psi_4$, is dominant in
the wave zone of a radiative gravitational field. On the other
hand, the Teukolsky \cite{teu} formalism for studying
gravitational radiation in a Kerr black hole is built using a
transverse frame, and the transverse components are the essential
variables. These and other similar facts focus attention on the
transverse frames and some {\it radiation scalars} have been
associated with them \cite{bee}.

The role played by the frames intrinsically associated with the
curvature tensor in the metric equivalence problem is well known
\cite{kra}. Likewise, both the transverse and Debever frames, can be
of interest in dealing with type I spacetimes. A detailed analysis
on the transformations leading to the standard canonical form of the
Weyl tensor for the different Petrov-Bel types has been given in
\cite{pol}, but some cases involving the solution of a quartic
equation have not been specified. In a recent paper \cite{rebru} a
general procedure is presented to obtain the transverse scalars in a
generic type I spacetime in which all the initial Weyl scalars are
non-vanishing. The method avoids solving the quartic equation but
the expressions are quite extended and complicated. Moreover this
aforementioned work does not present explicit expressions for the
transverse tetrad.

In this paper we present a general algorithm to determine {\it
all} the elements associated with {\it every} transverse frame in
a {\it generic} type I spacetime and starting from an {\it
arbitrary} initial tetrad. The procedure only uses a particular
solution of a cubic equation and it affords, for each of the three
principal planes: (i) the transverse scalars $\Psi_0^{T},
\Psi_2^{T}, \Psi_4^{T}$, (ii) the transverse base
$\{\textbf{\textit W}^{T}, \textbf{\textit U}^{T}, \textbf{\textit
V}^{T}\}$ of the self-dual bivector space, and (iii) the
transverse null tetrad $\{l^{T},n^{T},m^{T}, \bar{m}^{T}\}$.

It is worth remarking that obtaining the three principal transverse
frames in a type I spacetime can be suitable because, in dealing
with the equivalence problem of two metrics, it could be necessary
to compare a transverse frame of one metric with each of the three
transverse frames of the other. The algorithmic obtention of these
frames could also be a necessary mathematical tool for studying
gravitational radiation in numerical relativity as Beetle and Burko
have pointed out in \cite{bee}. Some recent alternative approaches
emphasize this aspect \cite{bee2} \cite{bee3}.

The results in this paper are based on a previous paper \cite{fsI}
which offers a covariant algorithm to determine the type I Weyl
canonical frames. This approach was further developed in
\cite{fms} where a complete algebraic analysis of the Weyl tensor
is presented. In this aforementioned paper {\it every} Weyl
geometric element (invariant scalars, principal directions or
principal planes) associated with {\it every} Petrov-Bel type is
determined in a covariant way. This means that, given a metric in
an arbitrary coordinate system or in an arbitrary tetrad, if we
know the Weyl eigenvalues, we can obtain all these Weyl geometric
elements without solving any other equation. The Weyl eigenvalues
are the roots of the (cubic) characteristic equation of the
self-dual Weyl endomorphism, whose coefficients are the symmetric
algebraic invariant scalars of the Weyl tensor. Thus, this
covariant approach could also be useful in looking for
longitudinal scalars and Debever frames, as well as, in dealing
with algebraically special Petrov-Bel types.

This work is organized as follows. In section 2 we summarize some
of the results in \cite{fsI} \cite{fms} and, in section 3 and 4,
we use them to obtain the elements associated with the transverse
frames. We will finish with a short discussion.

\section{\large THE COVARIANT PROCEDURE}

In an oriented spacetime $(V_4,g)$ of signature $(-+++)$ the
algebraic classification of the Weyl tensor $C$ can be obtained by
studying the linear map defined on the self-dual 2-form space by
the self-dual Weyl tensor ${\cal C}=\frac{1}{2} (C-i*C)$. The
restriction on this self-dual space of the metric of the 2--form
space $G$ is given by ${\cal G}=\frac{1}{2} (G-i \eta)$, where
$\eta$ is the spacetime volume element.  In terms of the complex
invariant scalars, $I= \frac12 \tr {\cal C}^{2}$, $J= \frac16 \tr
{\cal C}^{3}$, the characteristic equation takes the form:
\begin{equation} \label{eccar}
x^{3}-I x -2J =0
\end{equation}
and its roots are, for $k=0,1,2$,
\begin{equation}  \label{valp}
\alpha_{k+1} = \beta e^{\frac{2
\pi k}{3}\ci} + \frac{I}{3 \beta} e^{-\frac{2 \pi k}{3}\ci}\, , \qquad
\beta= \sqrt[3]{\left(J+\sqrt{J^2-I^3/27}\right)}.
\end{equation}
The Weyl tensor is Petrov-Bel type I if (\ref{eccar}) admits three
different roots $\{ \alpha_{i} \}$, which is equivalent to the
condition $27 J^{2} \neq I^{3}$. In this case an orthonormal frame
$\{\textit{\textbf W}_{i} \}$ exists which is built up with
eigenvectors of ${\cal C}$. The self-dual 2--forms
$\textit{\textbf W}_{i}$ are the {\it principal {\rm 2}-forms} of
the Weyl tensor \cite{bel}. Then, the self-dual Weyl tensor takes
the canonical expression
\begin{equation} \label{can1}
{\cal C}=- \sum_{i=1}^{3} \alpha_{i} \  \textit{\textbf W}_{i}
\otimes \textit{\textbf W}_{i}
\end{equation}

In \cite{fsI} we have determined the projection map on the
eigenspace associated with every eigenvalue $\alpha_i$ and,
consequently, we have acquired a covariant way to obtain the
principal 2-forms $\{\textit{\textbf W}_{i} \}$ in terms of the
Weyl tensor. More precisely, we can find in \cite{fsI} the
following result which we present here in a slightly different
version:

\begin{proposition} \label{prop1}
Let ${\cal C}$ be the self-dual Weyl tensor of a type I
space-time. The principal {\rm 2}-form $\textit{\textbf W}_{i}$
corresponding to the eigenvalue $\alpha_{i}$ may be obtained as
\begin{equation} \label{p2f}
\textit{\textbf W}_{i}=\frac{{\cal P}_{i}(\textit{\textbf
X})}{\sqrt{(I - 3 \alpha_i^2){\cal P}_{i}(\textit{\textbf
X},\textit{\textbf X})}}
\end{equation}
with $\ {\cal P}_{i} \equiv {\cal C}^{2}+\alpha_{i} {\cal
C}+(\alpha_{i}^2- I){\cal G}$, and where $\textit{\textbf X}$ is
an arbitrary self-dual 2-form such that ${\cal
P}_{i}(\textit{\textbf X}) \neq0$.
\end{proposition}

The principal 2-forms of a type I Weyl tensor are given by
(\ref{p2f}) and are determined up to sign and permutation. Thus,
we can consider 24 oriented eigen-frames $\{\textit{\textbf
W}_{i}\}$: for every permutation, the sign of two of them gives us
4 possibilities, and the third can be obtained as \cite{fsI}
$\textit{\textbf W}_{3} = \ci\sqrt{2}\textit{\textbf W}_{1} \times
\textit{\textbf W}_{2}$, where $\times$ stands for the contraction
of adjacent indexes, $(A \times B)_{\alpha \beta} = A_{\alpha}^{\
\lambda} B_{\lambda \beta}$.

The three principal 2--forms $\textit{\textbf W}_i$ determine six
Weyl principal 2--planes that cut in four orthogonal directions: one
{\it time-like Weyl principal direction} and three {\it space-like
Weyl principal directions} \cite{bel} \cite{fsI}, which have
associated the {\it Weyl canonical orthonormal frames} $\{
e_{\alpha} \}$. It has been shown in \cite{fsI} that an oriented and
orthochronous canonical orthonormal frame $\{ e_{\alpha} \}$
corresponds biunivocally to every oriented eigen-frame
$\{\textit{\textbf W}_{i} \}$. Its explicit expression is given in
the following proposition \cite{fsI}:

\begin{proposition} \label{prop2}
The Weyl canonical frames $\{ e_{\alpha} \}$ of a type I
space-time may be determined as
\begin{equation} \label{canf}
\hspace{-0.2cm} e_{0}=\frac{-P_{0} (x)}{\sqrt{P_{0}(x,x)}}\, ,
\quad P_0 \equiv -\frac{1}{2} \left(\frac{1}{2}g + \sum_{i=1}^{3}
\textit{\textbf W}_{i} \times \overline{\! \textit{\textbf W}}_{i}
\right)  ; \quad e_{i}=\sqrt{2} \textit{\textbf W}_{i}(e_0)
\end{equation}
where $\textit{\textbf W}_{i}$ are the principal 2-forms given in
Proposition \ref{prop1}, and where $x$ is an arbitrary
future-pointing vector and an overline stands for the complex
conjugate.
\end{proposition}

In the following sections we will see that the results in
propositions \ref{prop1} and \ref{prop2} enable us to obtain the
transverse scalars and the transverse frames starting from an
arbitrary frame. The covariant method presented in \cite{fsI}
could also be used in looking for the longitudinal scalars and
frames because the Debever directions of a type I spacetime have
also been obtained in \cite{fsI} explicitly. It is worth
mentioning that this study has been extended in \cite{fms} for an
arbitrary Petrov-Bel type in such a way that {\it given a
particular cubic root} $\beta$ of the expression (\ref{valp}), all
the Weyl geometric elements ({\it principal} 2--{\it forms}, {\it
principal and Debever directions} and {\it canonical frames}) can
be obtained without solving any other equation.

\section{\large PRINCIPAL TRANSVERSE FRAMES. TRANSVERSE SCALARS}

An arbitrary null tetrad $\{ l , k, m, \bar{m} \}$ has the
following associated null base $\{ \textit{\textbf W},
\textit{\textbf U}, \textit{\textbf V}\, \}$ of bivectors
\cite{cite1}:
\begin{equation}  \label{lU}
\textit{\textbf W} = \frac{1}{\sqrt{2}}[l \wedge
k + m \wedge \bar{m}], \quad  \textit{\textbf U} =
-\frac{1}{\sqrt{2}}l \wedge \bar{m} , \quad \textit{\textbf V} =
\frac{1}{\sqrt{2}} k \wedge m
\end{equation}

Let $\Psi_a = \Psi_0, \Psi_1, \Psi_2, \Psi_3, \Psi_4$ be the usual
components \cite{kra} of the Weyl tensor in the base of the
traceless double bivectors, i. e.:
$$\frac12\, {\cal C}= \Psi_0 \textit{\textbf U} \otimes
\textit{\textbf U} + \Psi_1 \textit{\textbf U}
\stackrel{\sim}{\otimes} \textit{\textbf W} + \Psi_2
(\textit{\textbf U} \stackrel{\sim}{\otimes} \textit{\textbf V} +
\textit{\textbf W} \otimes \textit{\textbf W} ) + \Psi_3
\textit{\textbf W} \stackrel{\sim}{\otimes} \textit{\textbf V} +
\Psi_4 \textit{\textbf V} \otimes \textit{\textbf V} $$ In terms
of the Weyl scalars $\Psi_a$, the invariants $I, J$ are
\cite{kra}:
\begin{eqnarray}
I & = & I[\Psi_{\alpha}] \equiv \Psi_0 \Psi_4 - 4 \Psi_1 \Psi_3 +
3 \Psi_2^2   \label{ipsi} \\
J & = & J[\Psi_{\alpha}] \equiv \left|
\begin{array}{ccc}
  \Psi_4 & \Psi_3 & \Psi_2 \\
  \Psi_3 & \Psi_2 & \Psi_1 \\
  \Psi_2 & \Psi_1 & \Psi_0 \\
\end{array}
\right|                     \label{jpsi}
\end{eqnarray}
Then, the Weyl eigenvalues can be computed by (\ref{valp}) where
$\beta$ stands for a particular non null cubic root of
(\ref{valp}) which, from now, we can suppose that it is given in
terms of $\Psi_a$ as a consequence of (\ref{ipsi}) and
(\ref{jpsi}):
\begin{equation} \label{betapsi}
\beta= \beta[\Psi_a] \equiv \sqrt[3]{\left(J+\sqrt{J^2-I^3/27}\right)}
\end{equation}

The Weyl {\it transverse bivector bases} $\{\textbf{\textit
W}^{T}, \textbf{\textit U}^{T}, \textbf{\textit V}^{T}\}$ are
those for which $\Psi^T_1 = \Psi^T_3 =0$. Thus, the Weyl tensor
can be written:
\begin{equation} \label{transverse}
\frac12 {\cal C} = \Psi_0^T \ \textbf{\textit U}^{T}\otimes
\textbf{\textit U}^{T} + \Psi_2^T \ (\textbf{\textit U}^{T}
\stackrel{\sim}{\otimes} \textbf{\textit V}^{T} + \textbf{\textit
W}^{T} \otimes \textbf{\textit W}^{T}) \ + \Psi_4^T \
\textbf{\textit V}^{T} \otimes \textbf{\textit V}^{T}
\end{equation}
Then, $\textbf{\textit U}^{T}$ and $\textbf{\textit V}^{T}$ can be
parameterized to satisfy $\Psi_0^T =\Psi_4^T $. If we compare
(\ref{transverse}) with the canonical expression (\ref{can1}) we
obtain that the tern $\{ \textbf{\textit W}^{T}, - \ci
(\textbf{\textit V}^{T} + \textbf{\textit U}^{T}), \textbf{\textit
V}^{T} - \textbf{\textit U}^{T} \}$ is an orthonormal eigenframe
of the Weyl tensor if $\{\textbf{\textit W}^{T}, \textbf{\textit
U}^{T}, \textbf{\textit V}^{T}\}$ is a transverse bivector base.
Conversely, if $\{\textit{\textbf W}_{i}\}$ is an orthonormal Weyl
eigenframe, for every choice of an eigenvalue $\alpha_i$, we can
take:
\begin{equation} \label{UV-W}
\textit{\textbf W}_i = \textit{\textbf W}_i \, , \qquad
 \textit{\textbf U}_i = \frac{\ci}{2}(\textit{\textbf W}_j + \ci
\textit{\textbf W}_k) \, , \qquad  \textit{\textbf V}_i =
\frac{\ci}{2}(\textit{\textbf W}_j - \ci \textit{\textbf W}_k) \,
,
\end{equation}
where $i,j,k$ take the different values of a cyclic permutation.
Then the tern $\{ \textit{\textbf W}_i, \textit{\textbf U}_i,
\textit{\textbf V}_i \}$ is an oriented bivector transverse frame
that we name {\it principal transverse bivector base}. Moreover,
for every $i$, the non null {\it principal transverse scalars}
$\Psi_a^{(i)}$ are given in terms of the eigenvalues of the Weyl
tensor by:
\begin{equation}  \label{tsvp}
\Psi_2^{(i)} =- \frac{1}{2} \alpha_i , \qquad  \Psi_0^{(i)} =
\Psi_4^{(i)} = \frac{\alpha_j - \alpha_k }{2}
\end{equation}
where $i,j,k$ take the different values of a cyclic permutation.

Consequently, three oriented {\it principal transverse bases} $\{
\textit{\textbf W}_i, \textit{\textbf U}_i, \textit{\textbf V}_i
\}$ exist basically (the others can be obtained by changing the
sign of two elements or by changing one sign and interchanging
$\textit{\textbf U}_i  \leftrightarrow \textit{\textbf V}_i$). For
each of these frames, we can give the corresponding {\it principal
transverse scalars} $\Psi_a^{(i)}$ by using (\ref{tsvp}). Indeed,
taking into account the expression of the eigenvalues (\ref{valp})
in terms of $I$, $J$ and $\beta$, we have:

\begin{theorem}  \label{theo1}
Let $\Psi_a$ be the components of a type I Weyl tensor in an
arbitrary frame. The principal transverse scalars are given by
(\ref{tsvp}), with:
\begin{eqnarray}
\alpha_1 & = & \alpha_1[\Psi_a] \equiv  \left(\beta +
\frac{I}{3\beta}\right)
\, , \label{alpha1} \\[1mm]
\alpha_2 & = & \alpha_2[\Psi_a] \equiv  - \frac{1}{2}
\left((1-{\rm i}\sqrt{3})\beta +
(1+{\rm i}\sqrt{3})\frac{I}{3\beta}\right) \, ,  \label{alpha2}  \\[1mm]
\alpha_3 & = & \alpha_3[\Psi_a] \equiv - \frac{1}{2} \left((1+{\rm
i}\sqrt{3})\beta + (1-{\rm i}\sqrt{3})\frac{I}{3\beta}\right) \,
\, .  \label{alpha3}
\end{eqnarray}
where $I \equiv I[\Psi_a]$, $J \equiv J[\Psi_a]$ and $\beta \equiv
\beta[\Psi_a]$ are given by (\ref{ipsi}), (\ref{jpsi}) and
(\ref{betapsi}), respectively.
\end{theorem}

If $\{ \textit{\textbf W}_i, \textit{\textbf U}_i, \textit{\textbf
V}_i \}$ is a principal transverse bivector base, then the tern:
\begin{equation}  \label{nptf}
\textbf{\textit W}^T = \textit{\textbf W}_i \, , \qquad
\textbf{\textit U}^T = z \textit{\textbf U}_i \, , \qquad
\textbf{\textit V}^T = z^{-1} \textit{\textbf V}_i
\end{equation}
where $z$ is a complex function, is a transverse bivector base
because it satisfies the transverse condition, $\Psi_1^{T} =
\Psi_3^{T} =0$. Moreover, the transverse components are:
\begin{equation} \label{nptc}
\Psi_2^{T} = \Psi_2^{(i)} \, , \qquad  \Psi_0^{T} = z^{-2}
\Psi_0^{(i)} \, , \qquad  \Psi_4^{T} = z^2 \Psi_4^{(i)} \, .
\end{equation}

It is worth pointing out that the transverse principal components
(given by (\ref{tsvp}) and theorem \ref{theo1}) are invariant
scalars, but for a generic (non necessarily principal) transverse
frame only the Coulomb component $\Psi_2^T$ is invariant.
Nevertheless, it follows from (\ref{tsvp}) and (\ref{nptc}) that
the product of the transverse components does not depend on the
complex Lorentz rotation $z$:
\begin{equation}
\xi^{(i)} \equiv \Psi_0^T \Psi_4^T = (\Psi_0^{(i)})^2 =
(\Psi_4^{(i)})^2
\end{equation}
The invariant scalars $\xi^{(i)}$ are not but the Beetle-Burko
radiation scalars which have been proposed in \cite{bee} as
containing information about the field gravitational radiation.

Our simple method to determine the transverse components starting
from an arbitrary frame presented in theorem \ref{theo1} improves
some previous results \cite{rebru} that offer quite complicated
expressions of the transverse components in a (not necessarily
principal) transverse frame.

The transformations leading from an initial configuration to the
transverse frames have been studied in \cite{pol}, but these
transformation have not been obtained explicitly in the more regular
cases. This problem is analyzed and solved in the following section
by offering explicit expressions for the transverse frames.

\section{\large OBTAINING TRANSVERSE FRAMES FROM AN ARBITRARY FRAME}

In order to determine the transverse frames starting from an
arbitrary frame, we begin by obtaining the principal 2--forms
$\{\textit{\textbf W}_{i} \}$ in terms of the initial bivector
base $\{ \textit{\textbf W}, \textit{\textbf U}, \textit{\textbf
V} \}$ and the initial Weyl scalars $\Psi_a$. We will use
proposition \ref{prop1} and, consequently, we must pick out a
self-dual 2-form $\textit{\textbf X}$. If we take $\textit{\textbf
X} = \textit{\textbf U}$, and we compute ${\cal
P}_i(\textit{\textbf U})$, where the projector ${\cal P}_i$ is
given in proposition \ref{prop1}, we obtain:

\begin{equation} \label{piu}
{\cal P}_i (\textit{\textbf U}) = A_i \ \textit{\textbf U} + B_i \
\textit{\textbf W} + C_i \textit{\textbf V}\, ,  \quad \qquad
{\cal P}_i (\textit{\textbf U},\textit{\textbf U}) = \frac12 C_i
\, ,
\end{equation}
where the scalars $A_i$, $B_i$ and $C_i$ are the following
functions of the Weyl components $\Psi_a$:
\begin{eqnarray}
A_i & = & A_i[\Psi_a] \equiv \Psi_0 \Psi_4 +  \Psi_2^2 - 2 \Psi_1
\Psi_3 + \alpha_i
\Psi_2 + (\alpha_i^2 - I)  \label{A} \\[2mm]
B_i & = & B_i[\Psi_a] \equiv \Psi_1 \Psi_4 - \Psi_2 \Psi_3 +
\Psi_3 \alpha_i \label{B} \\[2mm]
C_i & = & C_i[\Psi_a] \equiv 2 \Psi_2 \Psi_4 - 2 \Psi_3^2 +
\alpha_i \Psi_4  \label{C}
\end{eqnarray}
Then, if we take into account expression (\ref{p2f}) for the
principal 2--forms, we can state:
\begin{theorem}  \label{theo2}
Let $\Psi_a$ be the components of a type I Weyl tensor in a non
transverse bivector base $\{ \textit{\textbf W}, \textit{\textbf
U}, \textit{\textbf V} \}$. The (unitary) principal {\rm 2}--forms
$\{\textit{\textbf W}_{i} \}$ are given by:
\begin{equation} \label{uipsi}
\textit{\textbf W}_{i} = \frac{1}{\sqrt{D_i  C_i }}( A_i
\textit{\textbf U} + B_i \textit{\textbf W} + C_i \textit{\textbf
V}) \, ,
\end{equation}
where the scalars $A_i$, $B_i$ and $C_i$ are the functions of the
Weyl scalars $\Psi_a$ given in (\ref{A}), (\ref{B}) and (\ref{C}),
and
\begin{equation}
D_i = D_i[\psi_a] \equiv \frac12 (I - 3 \alpha_i^2) \, ,
\label{D}
\end{equation}
with $\alpha_i = \alpha_i[\Psi_a]$ given in theorem \ref{theo1}
and where $I \equiv I[\Psi_a]$, $J \equiv J[\Psi_a]$ and $\beta
\equiv \beta[\Psi_a]$ are given by (\ref{ipsi}), (\ref{jpsi}) and
(\ref{betapsi}), respectively.
\end{theorem}

It should be mentioned that if we start from a non transverse
frame as theorem \ref{theo2} states, the bivector $\textit{\textbf
U}$ can not be orthogonal to any principal 2--form
$\textit{\textbf W}_{i}$ and, consequently, the condition ${\cal
P}_i(\textit{\textbf U}) \not= 0$ required in proposition
\ref{prop1} holds.

Now, if we consider the relation (\ref{UV-W}) between a null base
and an orthonormal base in the bivector space, the determination
of the principal transverse bivector bases is a simple consequence
of theorem \ref{theo2}:

\begin{corollary}  \label{cor1}
Let $\Psi_a$ be the components of a type I Weyl tensor in an
arbitrary bivector base $\{ \textit{\textbf W}, \textit{\textbf
U}, \textit{\textbf V} \}$. The principal transverse bivector
bases $\{ \textit{\textbf W}_i, \textit{\textbf U}_i,
\textit{\textbf V}_i \}$ can be obtained as (\ref{UV-W}) where
$\{\textit{\textbf W}_{i} \}$ is given in theorem \ref{theo2}.
\end{corollary}
Theorem \ref{theo2} and corollary \ref{cor1} give us explicit
expressions for the three principal transverse bivector bases. The
non principal transverse bivector bases $\{\textbf{\textit W}^{T},
\textbf{\textit U}^{T}, \textbf{\textit V}^{T}\}$ can be obtained
from the principal ones as (\ref{nptf}) by considering arbitrary
values for the complex function $z$.

Once the transverse bivector bases $\{\textbf{\textit W}^{T},
\textbf{\textit U}^{T}, \textbf{\textit V}^{T}\}$ are known, we
can look for the transverse frames $\{l^{T},n^{T},m^{T},
\bar{m}^{T}\}$ associated with them by (\ref{lU}). To obtain them,
we could apply to the bivectors $\{\textbf{\textit W}^{T},
\textbf{\textit U}^{T}, \textbf{\textit V}^{T}\}$ the covariant
method to determine the principal directions of a 2--form
\cite{cf} (see also \cite{fsI} \cite{fms}), but here we opt by an
alternative procedure based on proposition \ref{prop2}: starting
from an arbitrary null tetrad  $\{ l , k, m, \bar{m} \}$ we will
obtain the Weyl orthonormal frame $\{e_{\alpha}\}$ and, from it,
we derive the null transverse frames $\{l^{T},k^{T},m^{T},
\bar{m}^{T}\}$.

From (\ref{uipsi}) we can obtain the $e_0$-projector $P_0$ given
in proposition \ref{prop2}. If we take $x=l$ and compute $P(l)$,
we have:
\begin{equation}
-4P(l) =  a l + b k + c  m + \bar{c} \bar{m} \, , \quad \quad
4P(l,l) = b  \label{Pl}
\end{equation}
where
\begin{eqnarray}
a & = & a[\Psi_a] \equiv 2 + \frac{c \bar{c}}{b}  \nonumber \\
b & = & b[\Psi_a] \equiv \sum{\frac{|C_i|}{|D_i|}}  \label{abc} \\
c & = & c[\Psi_a] \equiv  -\sum \frac{\bar{B}_i C_i } {|D_i| |C_i|
}  \nonumber
\end{eqnarray}

On the other hand, if we take into account the expression for the
$e_i$ given in proposition \ref{prop2}, we can state:

\begin{theorem}  \label{theo3}
Let $\Psi_a$ be the components of a type I Weyl tensor in a non
transverse null frame  $\{ l , k, m, \bar{m} \}$. The orthonormal
Weyl canonical frame can be obtained as:
\begin{equation}
e_0  = \frac{1}{2\sqrt{b}} \  ( a l + b k + c  m + \bar{c} \bar{m}
)  \, , \quad \qquad  e_i = \sqrt{2} \textit{\textbf W}_{i}(e_0)
\label{e0ei}
\end{equation}
where $\textit{\textbf W}_{i}$ are given in (\ref{uipsi}), $a, b ,
c$ are the functions of the Weyl scalars $\Psi_a$ given in
(\ref{abc}). The scalars $A_i$, $B_i$, $C_i$ and $D_i$ depend on
$\Psi_a$ as (\ref{A}), (\ref{B}), (\ref{C}) and (\ref{D}), where
$\alpha_i$ is given in theorem \ref{theo1} and $I$, $J$ and
$\beta$ are given by (\ref{ipsi}), (\ref{jpsi}) and
(\ref{betapsi}), respectively.
\end{theorem}

The principal transverse frames $\{ l_{(i)} , k_{(i)}, m_{(i)},
\bar{m}_{(i)} \}$ can be determined as a consequence of theorem
\ref{theo3}:
\begin{corollary}  \label{cor2}
Let $\Psi_a$ be the components of a type I Weyl tensor in an
arbitrary null frame  $\{ l , k, m, \bar{m} \}$. We can obtain the
principal transverse frames $\{ l_{(i)} , k_{(i)}, m_{(i)},
\bar{m}_{(i)} \}$ as:
\begin{equation}  \label{li}
l_{(i)} = \frac{1}{\sqrt{2}} (e_0 + e_i) \, , \quad k_{(i)} =
\frac{1}{\sqrt{2}} (e_0 - e_i) \, , \quad m_{(i)} =
\frac{1}{\sqrt{2}} (e_j + {\rm i} e_k)
\end{equation}
where $i,j,k$ take the different values of a cyclic permutation
and $e_{\alpha}$ are given in theorem \ref{theo3}.
\end{corollary}

From each one of the three principal transverse frames $\{ l_{(i)}
, k_{(i)}, m_{(i)}, \bar{m}_{(i)} \}$ determined in corollary
\ref{cor2} we can obtain the other (non principal) transverse null
frames  $\{l^{T},k^{T},m^{T}, \bar{m}^{T}\}$ by means a special
boost $\phi$ on the time-like plane $\{ l_{(i)} , k_{(i)} \}$ and
a rotation $\theta$ on the space-like plane $\{m_{(i)},
\bar{m}_{(i)} \}$:
\begin{equation}  \label{lT}
l^{T} = e^{\phi}l_{(i)} \, , \qquad k^{T}= e^{-\phi}k_{(i)} \, ,
\qquad m^{T}= e^{- \ci \theta}m_{(i)}  \, .
\end{equation}
The real functions $\phi$ and $\theta$ are related to the complex
Lorentz transformation $z$ in (\ref{nptf}) by $z= e^{\phi + \ci
\theta}$.

In this section we have obtained all the transverse frames of a
type I spacetime when we know the Weyl scalars in a non transverse
frame. But, if we initially have a transverse frame
$\{l^{T},n^{T},m^{T}, \bar{m}^{T}\}$, we could be interested in
knowing all the other ones. Let $\Psi_2^{T}, \Psi_0^{T},
\Psi_4^{T}$ be the initial transverse scalars. Then, as a
consequence of (\ref{nptc}), the complex Lorentz transformation
$z= e^{\phi + \ci \theta}$ leading to a principal transverse frame
$\{ l_{(i)} , k_{(i)}, m_{(i)}, \bar{m}_{(i)} \}$ by means of
(\ref{lT}) can be obtained as:
\begin{equation}
z = \left(\frac{\Psi_0^{T}}{\Psi_4^{T}}\right)^{1/4}
\end{equation}
Then, the transformations (\ref{li}) and their inverses allow us
to obtain the orthonormal canonical Weyl frame $\{e_{\alpha}\}$
and the other principal transverse frames.

\section{\large SUMMARY AND DISCUSSION}

In this paper we have presented a general algorithm to determine
{\it all} the elements associated with {\it every} transverse
frame in a {\it generic} type I spacetime and starting from an
{\it arbitrary} initial tetrad. Our procedure affords, for each of
the three principal planes: (i) the {\it principal transverse
scalars} $\Psi_2^{(i)}, \Psi_0^{(i)}, \Psi_4^{(i)}$ (theorem
\ref{theo1}) and the associated (non principal) transverse scalars
$\Psi_2^{T}, \Psi_0^{T}, \Psi_4^{T}$ (expression (\ref{nptc})),
(ii) the {\it principal bivector transverse base}
$\{\textbf{\textit W}_i, \textbf{\textit U}_i, \textbf{\textit
V}_i\}$ (theorem \ref{theo2} and corollary \ref{cor1}) and the
associated (non principal) transverse bases $\{\textbf{\textit
W}^{T}, \textbf{\textit U}^{T}, \textbf{\textit V}^{T}\}$
(expression (\ref{nptf})), and (iii) the {\it principal transverse
null tetrad} $\{l_{(i)},n_{(i)},m_{(i)}, \bar{m}_{(i)}\}$ (theorem
\ref{theo3} and corollary \ref{cor2}) and the associated (non
principal) transverse null tetrads $\{l^{T},n^{T},m^{T},
\bar{m}^{T}\}$ (expression (\ref{lT})).

We would like to state that the above results improve previous ones
on the same subject. We can quote an interesting work \cite{pol}
where a detailed analysis has been made of the transformations
leading to the standard canonical form of the Weyl tensor for the
different Petrov-Bel types. In the mentioned study the authors
present the transformations depending on the initial configuration
of the Weyl components $\Psi_a$, but "certain cases involving the
solution of a quartic equation have not been specified" (see Table 2
in \cite{pol}). It is worth mentioning, for example, the case of
type I spacetimes with non-vanishing initial transverse components.
This shortcoming has partially been overcome in a recent paper
\cite{rebru} where a procedure is presented "to arrive at a
transverse frame from the general case of all scalars non-zero" in a
given type I spacetime. The method avoids solving the quartic
equation but "the expressions for $\Psi_2$, $\Psi_0$ and $\Psi_4$
are quite complicated" according to the authors. Our paper overcomes
this shortcoming and we offer simple and clear expressions for the
transverse scalars. Moreover, we also obtain the transverse
components in the three principal transverse frames (theorem
\ref{theo1}) and in an arbitrary transverse frame (expression
(\ref{nptc})). On the other hand, obtaining the transformations that
were not studied in \cite{pol} is a problem that has not been
considered in \cite{rebru} either. This question is analyzed and
solved in this work by offering explicit expressions for all the
transverse frames.

\section*{\large ACKNOWLEDGMENTS}
The authors would like thank J.A. Morales for some useful
comments. This work has been supported by the Spanish Ministerio
de Ciencia y Tecnolog\'{\i}a, project AYA2003-08739-C02-02
(partially financed by FEDER funds).


\begin{thebibliography}{999}

\bibitem{sze} Szekeres, P. (1965). {\it J. Math. Phys.} {\bf 9}, 1387.
\bibitem{kra} Stephani, H., Kramer, D.,
 McCallum, M., Hoenselaers, C., and Herlt, E. (2003). {\it Exact
solutions of Einstein's field equations} (Cambridge University
Press, Cambridge).
\bibitem{bee} Beetle, C., and Burko, L.M. (2002) {\it Phys. Rew. Letters}
{\bf 89}, 271101.
\bibitem{sachs} Sachs, R. (1961). {\it Proc. R.
Soc. Lond.} {\bf A264}, 309.
\bibitem{teu} Teukolsky, S.A. (1972). {\it Phys. Rev. Lett.} {\bf 29}, 1114.
\bibitem{pol} Pollney, D., Skea, J.E.F., and d'Inverno, R.A. (2001)
{\it Class. Quantum Grav.} {\bf 17}, 643.
\bibitem{rebru} Re, V., Bruni, M., Matravers, D.R., and White, T. (2003).
{\it Gen. Rel. Grav.} {\bf 35}, 1351.
\bibitem{bee2} Beetle, C., Bruni, M., Burko, L.M., and Nerozzi., A.
(2004). Preprint gr-qc/0407012.
\bibitem{bee3} Nerozzi., A., Beetle, C., Bruni, M., Burko, L.M., and
Pollney, D. (2004). Preprint gr-qc/0407013.
\bibitem{fsI} Ferrando, J. J., and  S\'aez, J. A. (1997)
{\it Class. Quantum Grav.} {\bf 14}, 129.
\bibitem{fms} Ferrando, J. J., Morales, J. A., and S\'aez, J. A. (2001)
{\it Class. Quantum Grav.} {\bf 18}, 4939.
\bibitem{bel} Bel, L. (1962) {\it Cah. Phys.} {\bf 16}, 59.
(Engl. Transl. (2000) {\it Gen. Rel. Grav.} {\bf 32}, 2047.)
\bibitem{cite1} The notation that we use here is similar to that used in
reference \cite{kra}, but the self-dual bivectors $\{
\textit{\textbf W}, \textit{\textbf U}, \textit{\textbf V} \}$
differ from those in \cite{kra} by the factor $1/\sqrt{2}$, and
the self-dual Weyl tensor by the factor $1/2$.
\bibitem{cf} Coll, B. and Ferrando, J.J. (1988).
{\it Gen. Rel. Grav.} {\bf 20}, 51.


\end{thebibliography}
\end{document}